\documentclass[preprint,showpacs,superscriptaddress,groupedaddress,nofootinbib]{revtex4-2}
\usepackage{graphicx}
\usepackage{dcolumn}   
\usepackage{bm}   
\usepackage{amssymb}
\usepackage{color}

\usepackage{amsmath}
\usepackage{mathrsfs}
\usepackage{latexsym}
\usepackage{natbib}    
\usepackage[toc,page]{appendix}
\usepackage{braket}
\newcommand{\del}{\partial}

\begin{document}

\title{Exact Solutions with Asymptotically Flat Galactic Rotation Curves}

\author{Sandipan Sengupta}
\email{sandipan@phy.iitkgp.ac.in}
\affiliation{Department of Physics, Indian Institute of Technology Kharagpur, Kharagpur-721302, INDIA}

\begin{abstract}
We present a new set of spacetime solutions for spiral galaxies corresponding to flat rotation curves at large distances. These correspond to anisotropic pressure, which could be sourced by either  matter or geometry. 
As examples, we elucidate the galactic metrics corresponding to the non-ideal generalizations of `dust' and `radiation', respectively. Further, we find a dynamical realization of the Einstein cluster when the spatially averaged equation of state ($w$) happens to be non-zero but small. The correction to the Einsteinian bending angle for a light ray penetrating a large galactic halo is predicted to be: $\Big(\frac{2+3w}{1+3w}\Big) \frac{\pi v^2}{c^2}$, $v$ being the limiting rotational velocity. We infer a mild decline of the rotation curves ($v'(r)\sim\frac{1}{r^2}$) at large radii for sufficiently luminous galaxies, as also indicated by some specific observations.

\end{abstract}

\maketitle
\newpage


\section{Introduction}

The flatness of rotation curves far away from the luminous mass for spiral galaxies is empirically well-established \cite{rubin,*rubin1,roberts}. These observations are generally perceived as an indirect proof of the existence of either a new form (`dark') of matter or of an underlying gravity theory that is different from Einstein's. 

The search for a theoretical model of the hypothetical `dark matter' (DM) or a geometric theory of spacetime capable of explaining the flat profile of rotation curves at asymptotia have a long history \cite{bertone,salucci,einasto,nfw,servant,milgrom,*milgrom1,bek,mannheim,*mannheim1,harko,kar1,moffat,sobouti,
mendoza}. While the viewpoint that `dark matter' is an ideal fluid with zero or negligible pressure has been fairly popular owing to its simplicity \cite{nucamendi,barranco}, the possibility that it could correspond to anisotropic pressure has also received occasional attention \cite{bharadwaj,lake}. Among these earlier works, Ref.\cite{bharadwaj} reports a set of solutions under a number of rather special assumptions, that is, the radial and transverse equations of state are constant, the rotational velocities are exactly flat and the matter source are scalar fields.
 However, the important question as to what could be the generic class of galactic spacetime solutions (for spirals) corresponding to anisotropic pressure without all or at least some of these restrictive conditions has not been investigated so far.  Here we address this problem 
 and present some new results.

 For our purpose here, it is sufficient to assume that there exists an energy-momentum tensor whose origin could in principle be either an imperfect matter fluid or pure geometry. 
 However, to be specific, we restrict ourselves here to the confines of Einstein gravity with matter.

With this, we find a new class of galactic spacetime geometries as solutions to the field equations. These are parametrized by the limiting rotational velocity at the asymptotic distance far beyond luminous matter, and the spatially averaged equation of state parameter. Important observational implications involving the slope of the rotation curve and the gravitational deflection of light are also discussed. Furthermore, we explicitly demonstrate that there exists at least one class of extra dimensional theories \cite{sengupta,sengupta1}, namely, gravity theory with an extra dimension of zero proper length, which naturally admits some of these solutions as special cases. 

In the next section, we first formulate the dynamical equations for the spherically symmetric galactic spacetimes, before finding explicit solutions for galactic geometries. The resulting rotation curves and the properties of the stress-energy components are elucidated. Next we discuss
a few examples of the generic galactic geometries so obtained. This is followed by an analysis of the structure of the rotation curve
containing an explicit formula for its slope, as well as of the deflection of light passing through the galactic halo. We also elucidate an
intriguing connection between the solutions obtained here to extra dimensional gravity. The concluding section contains a summary
of results and a few relevant remarks. In Appendix A, we demonstrate how to determine explicitly the integration constants encoding
the solutions. The implications associated with the energy conditions are discussed in Appendix B.

\section{Construction of the galactic spacetime solution}

\subsection{Field equations}

The Einstein field equations are given by:
\begin{eqnarray}\label{EE}
R_{\alpha\beta}-\frac{1}{2}g_{\alpha\beta} R+\Lambda g_{\alpha\beta}=T_{\alpha\beta}
\end{eqnarray}
Here $T_{\alpha\beta}$ encodes the energy-momentum tensor associated with the `dark matter'. Alternatively, it could encode an effective energy momentum tensor of purely geometric origin, an example of which would be provided later.
Here our goal is to find spherically symmetric solutions to these equations for galactic spacetimes under generic conditions, without assuming any particular form of matter coupling or any specific equation of state. Also, we shall ignore the cosmological constant, a reasonable approximation in the context of galactic physics. 

Let us consider a spherically symmetric static model for the galactic halo, in line with the standard assumption ($ G=1=c$):
\begin{eqnarray*}
ds^2=-f(r)dt^2+g(r) dr^2+r^2(d\theta^2+\sin^2 \theta d\phi^2),
\end{eqnarray*}
The Einstein equations read:
\begin{eqnarray}\label{EFE}
&& \frac{1}{g}\Big[\frac{g'}{rg}+\frac{g-1}{r^2}\Big]=\rho,\nonumber\\
&& \frac{f'}{rfg}-\frac{1}{r^2}\Big(1-\frac{1}{g}\Big)=P_r,\nonumber\\
&& \frac{1}{2g}\Bigg[\frac{f''}{f}-\frac{f'}{2f}\Big(\frac{f'}{f}+\frac{g'}{g}\Big)+\frac{1}{r}\Big(\frac{f'}{f}-\frac{g'}{g}\Big)\Bigg]=P_\theta=P_\phi
\end{eqnarray}
Let us also note the trace equation, given by:
\begin{eqnarray}\label{trace1}
 \frac{1}{g}\Bigg[\frac{f''}{f}-\frac{f'}{2f}\left(\frac{f'}{f}+\frac{g'}{g}\right)+\frac{2}{r}\left(\frac{f'}{f}-\frac{g'}{g}\right)-\frac{2}{r^2}(g-1)\Bigg]~=~-\rho+P_r+2P_\theta 
\end{eqnarray}

The above system, with the variables $f(r),~g(r),~\rho(r),~P_r (r)$ and $P_\theta (r)$, is underdetermined. This reflects a generic property of Einstein gravity in presence of a nontrivial stress-energy tensor. Hence, we need two additional conditions to close the system. These are discussed next. 

\subsection{Solutions}

In vacuum Einstein theory, the unique spherically symmetric solution is the Schwarzschild spacetime, whose form is Newtonian: 
\begin{eqnarray}\label{newt}
f(r)g(r)=1,
\end{eqnarray} 
with $f(r)\rightarrow 1$ as $r\rightarrow \infty$. However, it is well-known that the vacuum Einstein equations do not contain galactic spacetimes as solutions, whose features are quite different. 
 To this end, let us first find a galactic analogue of the relation (\ref{newt}).
 

\subsection*{Exactly flat rotation curves:}

The circular velocity  of a massive test particle moving at the equatorial plane $\theta=\frac{\pi}{2}$ is given by: $v^2(r)=\frac{rf'(r)}{2f(r)}$ \cite{nucamendi}. For a rotation profile that is exactly flat at the outer radii of spirals, this implies: $v^2=const.\equiv \alpha$. This determines the function $f(r)$ upto an arbitrary  scale as:
\begin{eqnarray}\label{lambda}
f(r)=Ar^{2\alpha},
\end{eqnarray}
Since the limiting rotational velocity is different for each spiral galaxy, $\alpha$ emerges as a parameter in this solution. With this, we now proceed to find exact solutions with the property that the spatially averaged equation of state is constant:
\begin{eqnarray}\label{w}
\frac{\bar{P}}{\rho}\equiv \frac{\frac{1}{3}\Big(P_r+P_\theta+P_\phi\Big)}{\rho}=w
\end{eqnarray}
The trace equation (\ref{trace1}), upon using eqs.(\ref{lambda}), (\ref{w}) and the first equation in the set (\ref{EFE}), then leads to:
\begin{eqnarray}
(1+3w+\alpha)\frac{rg'(r)}{g^2(r)}-\frac{1+3w+2\alpha(1+\alpha)}{g(r)}+1+3w=0
\end{eqnarray}
This has the following solution:
\begin{eqnarray}\label{g-0}
g^{-1}(r)=\frac{1+3w}{1+3w+2\alpha(1+\alpha)}-Cr^{-\frac{1+3w+2\alpha(1+\alpha)}{1+3w+\alpha}}
\end{eqnarray}
To emphasize, the solution just found is only valid at the exactly flat rotation curve region. We note that the approximate solution found in Ref.\cite{nucamendi} under Newtonian conditions is a special case of the above family with $w=\frac{1-2\alpha^2}{3}$, which corresponds to an imperfect radiation fluid within our framework upto a small  ($o(\alpha^2)$) correction.
The rest of the equations in (\ref{EFE}) determine the appropriate density and pressure functions, thus solving all the components of the Einstein equations.  

Clearly the metrics defined by eqs.(\ref{lambda}) and (\ref{g-0}) are not asymptotically flat, in contrast with (\ref{newt}). The asymptotic limit of these solutions is  manifestly non-Newtonian (assuming a non-negative $w$):
\begin{eqnarray}\label{f-g}
f(r)g(r)\rightarrow A(1+\alpha+\alpha^2)r^{2\alpha}
\end{eqnarray}

\subsection*{Asymptotically flat rotation curves:}

Demanding that any galactic solution must be connected continuously to the unique large-distance limit (\ref{f-g}), we are naturally led to the galactic counterpart of the condition (\ref{newt}):
\begin{eqnarray}\label{fg-cond}
f(r)g(r)=\Big(\frac{r}{R}\Big)^{2\alpha},
\end{eqnarray}
where we have redefined the arbitrary constant in terms of the scale $R$. With this, we now proceed to find exact solutions with the property as earlier, that is, $w$ as defined in eq.(\ref{w}) is a constant.

Eqs.(\ref{fg-cond}) and (\ref{w}) along with the set (\ref{EFE}) now define a completely solvable system.
Inserting the first two into eq.(\ref{trace1}), we obtain:
\begin{eqnarray}
f''(r)+\frac{(3+3w-\alpha)}{r}f'(r)-\frac{2\alpha(1+3w)}{r^2}f(r)+\Big(\frac{1+3w}{r^2}\Big)\Bigg[f(r)-\Big(\frac{r}{R}\Big)^{2\alpha}\Bigg]=0
\end{eqnarray}
This leads to the following solution:
\begin{eqnarray}\label{fg}
f(r)~&=&~\left[\frac{1+3w}{(1-2\alpha)(1+3w)+2\alpha(2+\alpha+3w)}\right]\left(\frac{r}{R}\right)^{2\alpha}~-~ C_1 \left(\frac{r}{R}\right)^{\gamma_+}~+~C_2 \left(\frac{r}{R}\right)^{\gamma_-}\nonumber\\
~&=&~\Big(\frac{r}{R}\Big)^{2\alpha}g^{-1}(r)
\end{eqnarray}
where $\gamma_{\pm}=\frac{1}{2}
\left[-2-3w+\alpha\pm \Big((2+3w-\alpha)^2-4(1+3w)(1-2\alpha)\Big)^{\frac{1}{2}}\right]$ and $C_{1,2}$ are the integration constants. 

In the Newtonian limit $\alpha\rightarrow 0$, corresponding to a vanishing rotational velocity at asymptotia, eq.(\ref{fg}) reduces to:
\begin{eqnarray}\label{newt-lim}
f(r)= g^{-1}(r)\rightarrow 1-C_1\Bigg(\frac{R}{r}\Bigg)+C_2 \Bigg(\frac{R}{r}\Bigg)^{1+3w}
\end{eqnarray}
The second term above contains the Schwarzschild mass, whereas the last one depends on the matter content of the stress-energy tensor.

Finally, the solutions to all three independent components of the field equations (\ref{EFE}) read, respectively:
\begin{eqnarray}\label{rho}
8\pi\rho &~=~&\frac{1}{r^2}\Bigg[\frac{2\alpha(1+\alpha)}{(1-2\alpha)(1+3w)+2\alpha(2+\alpha+3w)}+  (1+\gamma_+ -2\alpha)C_1\left(\frac{r}{R}\right)^{\gamma_+ -2\alpha}\nonumber\\
&-& (1+\gamma_- -2\alpha)C_2 \left(\frac{r}{R}\right)^{\gamma_-- 2\alpha}\Bigg],\nonumber\\
8\pi P_r &~=~&\frac{1}{r^2}\Bigg[\frac{2\alpha(3w-\alpha)}{(1-2\alpha)(1+3w)+2\alpha(2+\alpha+3w)}- (1+\gamma_+)C_1\left(\frac{r}{R}\right)^{\gamma_+ -2\alpha}\nonumber\\
&+& (1+\gamma_-)C_2 \left(\frac{r}{R}\right)^{\gamma_- -2\alpha}\Bigg],\nonumber\\
8\pi P_\theta &~=~& \frac{1}{r^2}\Bigg[\frac{\alpha^2(1+3w)}{(1-2\alpha)(1+3w)+2\alpha(2+\alpha+3w)}\nonumber\\
&+&  \frac{1}{2}\Big((1+3w)(1+\gamma_+)-6w\alpha\Big)C_1\left(\frac{r}{R}\right)^{\gamma_+ -2\alpha} \nonumber\\
&-& \frac{1}{2}\Big((1+3w)(1+\gamma_-)-6w\alpha\Big)C_2 \left(\frac{r}{R}\right)^{\gamma_-- 2\alpha}\Bigg]~=~ 8\pi P_{\phi}
\end{eqnarray} 
Hence, the metric solution (\ref{fg}) corresponds to anisotropic  pressure: $P_r\neq P_\theta$.

Note that the apparent singularity of the metric at $r=0$ is not relevant, since it is only valid till an inner boundary at $r_0$ beyond which it should be continued to an interior metric associated with luminous matter.
 The associated mass function is given by:
\begin{eqnarray}\label{m}
M~&=&~4\pi \int^{R_0}_{r_0} dr' r'^2\rho(r')\nonumber\\
~&=&~\Big[\frac{\alpha(1+\alpha)}{(1-2\alpha)(1+3w)+2\alpha(2+\alpha+3w)}r~+~  \frac{C_1 R}{2}\left(\frac{r}{R}\right)^{\gamma_+ -2\alpha+1}\nonumber\\
~&-&~ \frac{C_2 R}{2} \left(\frac{r}{R}\right)^{\gamma_- -2\alpha+1}\Big]_{r_0}^{R_0}
\end{eqnarray}
where $r_0$ and $R_0$ are the inner and outer boundaries of the metric, respectively.

\subsection{Circular velocity}


Let us note that the condition (\ref{fg-cond}) alone does not determine the properties of the rotation curve $v(r)$. To investigate whether the above solutions are relevant for spiral galaxies, let us consider the resulting  circular velocity:
\begin{eqnarray}\label{vc}
v^2(r)=\frac{\frac{\alpha(1+3w)}{(1-2\alpha)(1+3w)+2\alpha(2+\alpha+3w)}- \frac{1}{2}  C_1 \gamma_+\left(\frac{r}{R}\right)^{\gamma_+ -2\alpha}+\frac{1}{2} C_2 \gamma_-\left(\frac{r}{R}\right)^{\gamma_- - 2\alpha}}{\frac{1+3w}{(1-2\alpha)(1+3w)+2\alpha(2+\alpha+3w)} -C_1 \left(\frac{r}{R}\right)^{\gamma_+ - 2\alpha}+C_2 \left(\frac{r}{R}\right)^{\gamma_-- 2\alpha}}
\end{eqnarray}
 Using the expressions for $\gamma_\pm$, the above corresponds to a flat rotation curve ($v^2\rightarrow \alpha$) as $r\rightarrow \infty$, as long as we have:  
 \begin{eqnarray}
 w>-\frac{1}{3},
\end{eqnarray}  
reflecting a theoretical lower bound on the `dark matter' within this framework. 
However, here we shall only consider the solutions with $w\geq 0$. Given that typically $\alpha\sim$ $10^{-6}$ to $10^{-5}$ for spirals, it could be treated as a small parameter.

\section{Examples}

\subsection{Non-ideal `dust':}

For a vanishing averaged pressure $\bar{P}$, we have a solution where the `dark matter' content represents a non-ideal generalization of the dust: $w=0$. The solution for geometry is given by the metric function as:
\begin{eqnarray}\label{w0}
f(r)=\frac{1}{(1+2\alpha+2\alpha^2)}\Big(\frac{r}{R}\Big)^{2\alpha}-C_1 \Big(\frac{r}{R}\Big)^{\frac{1}{2}\Big[-2+\alpha+\sqrt{\alpha(4+\alpha)}\Big]}+C_2 \Big(\frac{r}{R}\Big)^{\frac{1}{2}\Big[-2+\alpha-\sqrt{\alpha(4+\alpha)}\Big]}
\end{eqnarray}
The density and pressure components read:
\begin{eqnarray}\label{pw0}
8\pi\rho &~=~&\frac{1}{r^2}\Bigg[\frac{2\alpha(1+\alpha)}{1+2\alpha+2\alpha^2}-\frac{C_1}{2}(3\alpha-\sqrt{\alpha(4+\alpha)})\left(\frac{r}{R}\right)^{-1-\alpha+\frac{1}{2}\sqrt{\alpha(4+\alpha)}}\nonumber\\
&+& \frac{C_2}{2}(3\alpha+\sqrt{\alpha(4+\alpha)})\left(\frac{r}{R}\right)^{-1-\alpha-\frac{1}{2}\sqrt{\alpha(4+\alpha)}}\Bigg],\nonumber\\
8\pi P_r &~=~&\frac{1}{r^2}\Bigg[-\frac{2\alpha^2}{1+2\alpha+2\alpha^2}-\frac{C_1}{2}(\alpha+\sqrt{\alpha(4+\alpha)})\left(\frac{r}{R}\right)^{-1-\alpha+\frac{1}{2}\sqrt{\alpha(4+\alpha)}}\nonumber\\
&+& \frac{C_2}{2}(\alpha-\sqrt{\alpha(4+\alpha)})\left(\frac{r}{R}\right)^{-1-\alpha-\frac{1}{2}\sqrt{\alpha(4+\alpha)}}\Bigg],\nonumber\\
8\pi P_\theta &~=~&\frac{1}{r^2}\Bigg[\frac{\alpha^2}{(1+2\alpha+2\alpha^2)}+\frac{C_1}{4}(\alpha+\sqrt{\alpha(4+\alpha)})\left(\frac{r}{R}\right)^{-1-\alpha+\frac{1}{2}\sqrt{\alpha(4+\alpha)}}\nonumber\\
&-& \frac{C_2}{4}(\alpha-\sqrt{\alpha(4+\alpha)})\left(\frac{r}{R}\right)^{-1-\alpha-\frac{1}{2}\sqrt{\alpha(4+\alpha)}}\Bigg]
\end{eqnarray}
Given that $\alpha<<1$, the first term dominates the expression of density above as $r\rightarrow \infty$. This precisely reflects the behaviour underlying a flat rotation curve. Note that in this asymptotic limit, $\frac{P_r}{\rho}\rightarrow -\alpha,~\frac{P_\theta}{\rho}\rightarrow \frac{\alpha}{2}$. 
The associated mass function is obtained from eq.(\ref{m}) as:
\begin{eqnarray*}
M=\Bigg[\frac{\alpha(1+\alpha)r}{1+2\alpha+2\alpha^2}+\frac{C_1 }{2}\Big(\frac{r}{R}\Big)^{\frac{1}{2}(-3\alpha+\sqrt{\alpha(4+\alpha)})}-\frac{C_2 }{2}\Big(\frac{r}{R}\Big)^{\frac{1}{2}(-3\alpha-\sqrt{\alpha(4+\alpha)})}\Bigg]_{r_0}^{R_0}
\end{eqnarray*}

\subsection{Non-ideal `radiation':}

Let us consider the case $w=\frac{1}{3}$, which could be seen as an anisotropic generalization of radiation matter. The associated metric reads: 
\begin{eqnarray}\label{f-radn}
f(r)=\frac{1}{(1+\alpha+\alpha^2)}\Big(\frac{r}{R}\Big)^{2\alpha}-C_1 \left(\frac{r}{R}\right)^{\frac{1}{2}\Big[-3+\alpha+\sqrt{1+10\alpha+\alpha^2}\Big]}+C_2 \left(\frac{r}{R}\right)^{\frac{1}{2}\Big[-3+\alpha-\sqrt{1+10\alpha+\alpha^2}\Big]}
\end{eqnarray}
The stress-energy tensor components are displayed below:
\begin{eqnarray}
8\pi\rho &~=~&\frac{1}{r^2}\Bigg[\frac{\alpha(1+\alpha)}{(1+\alpha+\alpha^2)}+\frac{C_1}{2}(-1-3\alpha+\sqrt{1+10\alpha+\alpha^2})\left(\frac{r}{R}\right)^{\frac{1}{2}(-3-3\alpha+\sqrt{1+10\alpha+\alpha^2})}\nonumber\\
&+& \frac{C_2}{2}(1+3\alpha+\sqrt{1+10\alpha+\alpha^2})\left(\frac{r}{R}\right)^{\frac{1}{2}(-3-3\alpha-\sqrt{1+10\alpha+\alpha^2})}\Bigg]\nonumber\\
8\pi P_r &~=~&\frac{1}{r^2}\Bigg[\frac{\alpha(1-\alpha)}{(1+\alpha+\alpha^2)}+\frac{C_1}{2}(1-\alpha-\sqrt{1+10\alpha+\alpha^2})\left(\frac{r}{R}\right)^{\frac{1}{2}(-3-3\alpha+\sqrt{1+10\alpha+\alpha^2})}\nonumber\\
&+& \frac{C_2}{2}(-1+\alpha-\sqrt{1+10\alpha+\alpha^2})\left(\frac{r}{R}\right)^{\frac{1}{2}(-3-3\alpha-\sqrt{1+10\alpha+\alpha^2})}\Bigg]\nonumber\\
8\pi P_\theta &~=~&\frac{1}{r^2}\Bigg[\frac{\alpha^2}{(1+\alpha+\alpha^2)}+\frac{C_1}{2}(-1-3\alpha+\sqrt{1+10\alpha+\alpha^2})\left(\frac{r}{R}\right)^{\frac{1}{2}(-3-3\alpha+\sqrt{1+10\alpha+\alpha^2})}\nonumber\\
&+& \frac{C_2}{2}(1+3\alpha+\sqrt{1+10\alpha+\alpha^2})\left(\frac{r}{R}\right)^{\frac{1}{2}(-3-3\alpha-\sqrt{1+10\alpha+\alpha^2})}\Bigg]
\end{eqnarray}
For a large distance from the galactic centre, $\frac{P_r}{\rho}\rightarrow 1,~\frac{P_\theta}{\rho}\rightarrow \alpha $. Hence, unlike the `dust' solution, here the radial EOS is not small.
 The mass function (\ref{m}) now reads:
\begin{eqnarray*}
M=\Bigg[\frac{\alpha(1+\alpha)r}{2(1+\alpha+\alpha^2)}+\frac{C_1}{2}\Big(\frac{r}{R}\Big)^{\frac{1}{2}(-1-3\alpha+\sqrt{1+10\alpha+\alpha^2})}-\frac{C_2 }{2}\Big(\frac{r}{R}\Big)^{\frac{1}{2}(-1-3\alpha-\sqrt{1+10\alpha+\alpha^2})}\Bigg]_{r_0}^{R_0}~.
\end{eqnarray*}

\subsection{Einstein cluster:}

Here we consider the special class: $w=\frac{\alpha}{3}$. The resulting geometry is given by:
\begin{eqnarray}
f(r)~=~\frac{1}{(1+2\alpha)}\Big(\frac{r}{R}\Big)^{2\alpha}-C_1\left(\frac{r}{R}\right)^{-1+\sqrt{\alpha(1+2\alpha)}}+ C_2\left(\frac{r}{R}\right)^{-1-\sqrt{\alpha(1+2\alpha)}}
\end{eqnarray}
From eq.(\ref{rho}), we find:
\begin{eqnarray}
8\pi\rho &~=~&\frac{1}{r^2}\Bigg[\frac{2\alpha}{1+2\alpha}+C_1\Big(-2\alpha+\sqrt{\alpha(1+2\alpha)}\Big)\left(\frac{r}{R}\right)^{-1-2\alpha+\sqrt{\alpha(1+2\alpha)}}\nonumber\\
&+& C_2\Big(2\alpha+\sqrt{\alpha(1+2\alpha)}\Big)\left(\frac{r}{R}\right)^{-1-2\alpha-\sqrt{\alpha(1+2\alpha)}}\Bigg]
,\nonumber\\
8\pi P_r &~=~& \frac{1}{r^2}\Bigg[-C_1\sqrt{\alpha(1+2\alpha)}\left(\frac{r}{R}\right)^{-1-2\alpha+\sqrt{\alpha(1+2\alpha)}}+C_2\sqrt{\alpha(1+2\alpha)}
\left(\frac{r}{R}\right)^{-1-2\alpha-\sqrt{\alpha(1+2\alpha)}}\Bigg],\nonumber\\
8\pi P_\theta &~=~& \frac{1}{r^2}\Bigg[\frac{\alpha^2}{1+2\alpha}+\frac{C_1}{2}\Big((1+\alpha)\sqrt{\alpha(1+2\alpha)}-2\alpha^2)\left(\frac{r}{R}\right)^{-1-2\alpha+\sqrt{\alpha(1+2\alpha)}}\nonumber\\
&+&
\frac{C_2}{2}\Big((1+\alpha)\sqrt{\alpha(1+2\alpha)}+2\alpha^2)
\left(\frac{r}{R}\right)^{-1-2\alpha-\sqrt{\alpha(1+2\alpha)}}\Bigg]
\end{eqnarray}
Clearly, the radial pressure $P_r$ vanishes, whereas $P_\theta\rightarrow \frac{\alpha^2}{8\pi(1+2\alpha)r^2}$, as $r\rightarrow \infty$. This reflects a dynamical realization of the Einstein cluster within our framework. The transverse EOS in this limit becomes: $\frac{P_\theta}{\rho}\rightarrow \frac{\alpha}{2}$.

The mass reads:
\begin{eqnarray}
M=\Bigg[\frac{\alpha r}{(1+2\alpha)}+\frac{C_1}{2}\left(\frac{r}{R}\right)^{-2\alpha+\sqrt{\alpha(1+2\alpha)}}- \frac{C_2}{2}\left(\frac{r}{R}\right)^{-2\alpha-\sqrt{\alpha(1+2\alpha)}}\Bigg]_{r_0}^{R_0}~.
\end{eqnarray}

\section{Slope of rotation curve:}

Here we briefly discuss the slope of the rotation curves, which is important from an observational viewpoint.
We consider the two cases of a small and large $w$ parameter separately below. 


For a given $w$, the integration constants $C_{1,2}$ could be provided a physical interpretation through the Newtonian limit (\ref{newt-lim}), in a manner similar to Einsteinian gravity. 
For $w\approx 0$, the limit of the solution (\ref{fg}) implies:
\begin{eqnarray}
f(r)\approx 1-\frac{(C_1-C_2)R}{r}
\end{eqnarray}
Thus, $C_1-C_2\equiv 2m_B>0$ reflects the baryonic mass in the Newtonian limit. Using the expression (\ref{vc}) for the circular velocity, we obtain the leading order contribution as: 
\begin{eqnarray}\label{v^2}
\frac{dv(r)}{dr}\approx -\frac{m_B}{2\sqrt{\alpha}r^2}
\end{eqnarray}
In the above, we have used the weak-field approximation ($|C_{1,2}|<<1$), which is known to be a reasonable assumption for galactic potentials.
Thus, the rotation curve $v(r)$ in this case is predicted to exhibit a gentle decline along its approach to asymptotic distances.


For the other possibility where $w$ is positive but not quite small ($w>>\alpha$),
the baryonic mass  could be identified from the Newtonian limit (\ref{newt-lim}) as: $C_1 R\equiv 2m_B$. A concrete example is the case of non-ideal radiation ($w=\frac{1}{3}$) elucidated earlier:
\begin{eqnarray}
f(r)\approx 1-\frac{C_1 R}{r}+\frac{C_2 R^2}{r^{2}}
\end{eqnarray}
In the above, the coefficient $C_2$ in the second term resembles a `charge'.
Next, using the expression of $v(r)$ as earlier, the leading order contribution to the slope is obtained as:
\begin{eqnarray}\label{slope}
\frac{dv(r)}{dr}\approx -\frac{C_1}{4\sqrt{\alpha}r^2}+\frac{C_2(1+3w)^2}{4\sqrt{\alpha}r^{-2+3w}}
\end{eqnarray}
Clearly, at large distances, the first term dominates the second for sufficiently luminous or baryon dominated galaxies, leading to the following limiting formula for the slope:
\begin{eqnarray}
\frac{dv(r)}{dr}\approx -\frac{m_B}{2\sqrt{\alpha}r^2}
\end{eqnarray}
which is the same as eq.(\ref{v^2}), encoding a slight decline at large radii. 

A mild decline of the rotation curve (for fairly luminous galaxies) far away from the normal matter as predicted above is indeed what has been typically observed so far, for instance in the case of Milky way \cite{eilers,wang}. The fact that $v'(r)\rightarrow 0$ at asymptotic distances is a reflection of the asymptotic flatness of the rotation curves. Let us note though that for sufficiently low brightness, the slope could be positive in principle provided $C_2>0$ and the second term in eq.(\ref{slope}) dominates the first. 
%
 

\section{Gravitational deflection of light within a halo}

Let us consider the null geodesic equation obeyed by a light ray at the equatorial plane ($\theta=\frac{\pi}{2}$) for a large halo. The total angular distance covered in going from the radial distance ($r_0$) at closest approach from the centre of the halo to the observer at infinity is given by \cite{nucamendi}:
\begin{eqnarray}\label{Delta}
\Delta \phi&=&\int_{r_0}^\infty dr~\frac{d\phi}{dr}=\int_{r_0}^\infty \frac{dr}{r}~\left[\frac{E^2}{L^2}\frac{r^2}{fg}-\frac{1}{g}\right]^{-\frac{1}{2}}
\end{eqnarray}
where $E,L$ are the conserved quantities associated with the $t$ and $\phi$ motions, respectively. These satisfy the relation $\frac{E^2}{L^2}=\frac{f(r_0)}{r_0^2}$, $r_0$ being the distance at closest approach to the galactic centre within the halo spacetime. The total angle of deflection of the ray is then given by: 
\begin{eqnarray}\label{delta}
\delta=|2\Delta \phi-\pi|
\end{eqnarray}

Next, we analyze two possible cases (assuming $w\geq 0$) separately.

\subsection{Small $w$ parameter}

Let us first consider the possibility where $w<<1$, which includes the case $w=0$. Here we shall  use the identification of the baryon mass at the interior as: $\frac{1}{2}(C_1-C_2)R=m_B$, based on the Newtonian limit (\ref{newt-lim}).

Using the identity:
\begin{eqnarray}\label{id}
\frac{E^2}{L^2}\frac{r^2}{fg}-\frac{1}{g}&=&\frac{1+3w}{(1-2\alpha)(1+3w)+2\alpha(2+\alpha+3w)}\left[\left(\frac{r}{r_0}\right)^{2-2\alpha}-1\right]\times\nonumber\\
\Bigg[1&-&
\frac{(1-2\alpha)(1+3w)+2\alpha(2+\alpha+3w))}{1+3w}C_1 \left(\frac{r}{R}\right)^{\gamma_+}\frac{\left(\frac{r}{r_0}\right)^{2-2\alpha-\gamma_+}-1}{\left(\frac{r}{r_0}\right)^{2-2\alpha}-1}\nonumber\\
&+& \frac{(1-2\alpha)(1+3w)+2\alpha(2+\alpha+3w))}{1+3w}C_2 \left(\frac{r}{R}\right)^{\gamma_-}\frac{\left(\frac{r}{r_0}\right)^{2-2\alpha-\gamma_-}-1}{\left(\frac{r}{r_0}\right)^{2-2\alpha}-1}\Bigg],
\end{eqnarray}
the expression for $\Delta \phi$ for the passage of the light ray within the halo reads:
\begin{eqnarray}
\Delta \phi &\approx & \int_{r_0}^{\infty} \frac{dr}{r}~\Big[\frac{(1-2\alpha)(1+3w)+2\alpha(2+\alpha+3w)}{1+3w}\Big]^{\frac{1}{2}}\left[\left(\frac{r}{r_0}\right)^{2-2\alpha}-1\right]^{-\frac{1}{2}}\nonumber\\
&\times & \Bigg[1-
 \frac{(1-2\alpha)(1+3w)+2\alpha(2+\alpha+3w)}{2(1+3w)}\Bigg(C_1 \left(\frac{r}{R}\right)^{\gamma_+}\frac{\left(\frac{r}{r_0}\right)^{2-2\alpha-\gamma_+}-1}{\left(\frac{r}{r_0}\right)^{2-2\alpha}-1}\nonumber\\
 &&~+~C_2 \left(\frac{r}{R}\right)^{\gamma_-}\frac{\left(\frac{r}{r_0}\right)^{2-2\alpha-\gamma_-}-1}{\left(\frac{r}{r_0}\right)^{2-2\alpha}-1}\Bigg)\Bigg]\nonumber\\
&&\approx\Bigg[\frac{(1-2\alpha)(1+3w)+2\alpha(2+\alpha+3w)}{(1+3w)}\Bigg]^{\frac{1}{2}} \Bigg[\frac{\tan^{-1}\left(\left(\frac{r}{r_0}\right)^{2-2\alpha}-1\right)^{\frac{1}{2}}}{1-\alpha}\Bigg]_{r_0}^{\infty}\nonumber\\
&&-~\Bigg[\frac{(1-2\alpha)(1+3w)+2\alpha(2+\alpha+3w)}{(1+3w)}\Bigg]^{\frac{3}{2}}\frac{m_B}{r_0} \Bigg[\frac{\frac{2r}{r_0}+1}{\frac{r}{r_0}\Big(\frac{r}{r_0}+1\Big)}\Bigg]_{r_0}^{\infty}\nonumber\\
&&=~\Bigg[\frac{(1-2\alpha)(1+3w)+2\alpha(2+\alpha+3w)}{(1+3w)}\Bigg]^{\frac{1}{2}}\frac{\pi}{2(1-\alpha)}\nonumber\\
&&+~2\Bigg[\frac{(1-2\alpha)(1+3w)+2\alpha(2+\alpha+3w)}{(1+3w)}\Bigg]^{\frac{3}{2}}\frac{m_B}{r_0}
\end{eqnarray}
In writing the first line above, we have used the weak-field approximation for terms involving $C_{1,2}$.
Hence, we find the leading contribution (ignoring the small corrections in $\alpha,w$ beyond linear order) to the total deflection angle (\ref{delta}) as:
\begin{eqnarray}\label{defl}
\delta \approx \frac{4 m_B}{r_0}+2\pi\alpha+\frac{12 \alpha m_B}{r_0}+\frac{5}{2 }\pi \alpha^2
\end{eqnarray}
The first term above is the well-known Einstein bending, whereas the second term reflects the leading correction to it. 

Note that the above analysis applies to both the non-ideal `dust' ($w=0$) and the Einstein cluster ($w=\frac{\alpha}{3}$) solutions. The deflection angle for the Einstein cluster differs from the first only in $o(\alpha^2)$ corrections, which are too small though to be practically relevant.

The leading correction in the first case reproduces the nonbaryonic contribution ($ 2\pi \alpha$) predicted for a singular isothermal halo within the `cold dark matter' model (ideal dust fluid) \cite{blandford}. However, the crucial difference between the non-ideal and ideal dust is encoded in the density profiles, as is manifest from the earlier discussion. Further, the subleading corrections in (\ref{defl}) depend on $\frac{m_B}{r_0}$. This is in contrast with the singular CDM halo, where the subdominant corrections depend only on $\alpha$: $\Delta\delta\approx \frac{11}{2}\pi\alpha^2$.

\subsection{Large $w$ parameter}

Here we consider the case where $w\lesssim 1$ and hence $w>>\alpha$. Note that in this case, $\gamma_+\approx -1$ and $\gamma_-\approx -1-3w$ upto small corrections. This implies that the $C_2$ dependent term is subleading compared to the $C_1$-dependent term in the weak-field limit, and may be ignored. In this case, the baryon mass is identified to $C_1$ from the Newtonian limit as: $m_B=\frac{C_1 R}{2}$. Proceeding as earlier, we find:
\begin{eqnarray}
\Delta \phi &\approx & \int_{r_0}^{\infty} \frac{dr}{r}~\Big[\frac{(1-2\alpha)(1+3w)+2\alpha(2+\alpha+3w)}{1+3w}\Big]^{\frac{1}{2}}\left[\left(\frac{r}{r_0}\right)^{2-2\alpha}-1\right]^{-\frac{1}{2}}\nonumber\\
&\times & \Bigg[1-
 C_1\frac{(1-2\alpha)(1+3w)+2\alpha(2+\alpha+3w)}{2(1+3w)} \left(\frac{r}{R}\right)^{\gamma_+}\frac{\left(\frac{r}{r_0}\right)^2+\frac{r}{r_0}+1}{\frac{r}{r_0}+1}\Bigg]\nonumber\\
 &&\approx\Bigg[\frac{(1-2\alpha)(1+3w)+2\alpha(2+\alpha+3w)}{(1+3w)}\Bigg]^{\frac{1}{2}} \Bigg[\frac{\tan^{-1}\left(\left(\frac{r}{r_0}\right)^{2-2\alpha}-1\right)^{\frac{1}{2}}}{1-\alpha}\Bigg]_{r_0}^{\infty}\nonumber\\
&&-~\Bigg[\frac{(1-2\alpha)(1+3w)+2\alpha(2+\alpha+3w)}{(1+3w)}\Bigg]^{\frac{3}{2}}\frac{m_B}{r_0} \Bigg[\frac{\frac{2r}{r_0}+1}{\frac{r}{r_0}\Big(\frac{r}{r_0}+1\Big)}\Bigg]_{r_0}^{\infty}\nonumber\\
&&\approx \Bigg[1+\frac{(2+3w)\alpha}{1+3w}\Bigg]\frac{\pi}{2}+\frac{2m_B}{r_0}
\end{eqnarray}
where in the last line we have ignored the higher order corrections in $\alpha$ and $\frac{m_B}{r_0}$. Hence, the the deflection angle finally reads:
\begin{eqnarray}\label{Delta}
\delta=\frac{4m_B}{r_0}+\Big(\frac{2+3w}{1+3w}\Big)\pi\alpha
\end{eqnarray}
The second term above encodes the dominant correction to the Einstein bending.
For instance, for the special case of non-ideal radiation, the correction reads: $\Delta\delta=\frac{3}{2}\pi \alpha$, which is lesser than the non-ideal dust bending corresponding to $w\rightarrow 0$.

To emphasize, the results in ($\ref{defl}$) and ($\ref{Delta}$) make the formulation here amenable to lensing observations, and also to comparisons with various `dark matter' scenarios based on different density profiles or its geometric alternatives (e.g. see \cite{harko1,*pal}).

\section{Connection to extra dimensional gravity}

At the beginning, we mentioned the alternative possibility that the energy-momentum tensor $T_{\alpha\beta}$ in (\ref{EE}) could be generated by purely geometric effects under certain special circumstances. Here we explicitly show that that is indeed the case, where a geometry (\ref{fg}) for a given $w$ with flat rotation curves at large distances could arise as an exact solution to a gravity theory defined in higher than four spacetime dimensions. Here we focus on a 5D gravity theory where the fifth dimension exhibits a vanishing proper length, a formulation presented in detail in Ref.\cite{sengupta}.

The five-dimensional Lagrangian density describing the gravity theory is given by:
\begin{eqnarray}\label{L5}
{\cal L}(\hat{e},\hat{w})=\epsilon^{\mu\nu\alpha\beta\rho} \epsilon_{IJKLM}\hat{e}_{\mu}^{I}\hat{e}_{\nu}^{J}\hat{e}_{\alpha}^{K} \hat{R}_{\beta\rho}^{~~LM}(\hat{w}),
\end{eqnarray}
$\hat{e}_\mu^I$ and $\hat{w}_\mu^{~IJ}$ are the $SO(4,1)$ ($SO(3,2)$) vielbein and gauge connection, respectively. In this first order form, both are regarded as independent fields. We have chosen the five-dimensional Planck-length as unity and $\hat{R}_{\beta\rho}^{~~LM}(\hat{w})=\del_{[\beta} \hat{w}_{\rho]}^{~LM}+\hat{w}_{[\beta}^{~LK}\hat{w}_{\rho]K}^{~~~M}$ is the field-strength. 
A variation of (\ref{L5}) with respect to these leads to the following field equations in vacuum, respectively: 
\begin{eqnarray}
\delta \hat{w}&:&~~\epsilon^{\mu\nu\alpha\beta\rho} \epsilon_{IJKLM}\hat{e}_{\mu}^{I}\hat{e}_{\nu}^{J} \hat{D}_{\alpha}(\hat{w})\hat{e}_{\beta}^{K}=0,\nonumber\\
\delta \hat{e}&:&~~\epsilon^{\mu\nu\alpha\beta\rho} \epsilon_{IJKLM}\hat{e}_{\mu}^{I} \hat{e}_{\nu}^{J} \hat{R}_{\alpha\beta}^{~KL}(\hat{w})=0\label{eom}
\end{eqnarray}
Here $\hat{D}_\mu(\hat{w})$ denotes the gauge-covariant derivative with respect to $\hat{w}_\mu^{~IJ}$. Clearly, the Lagrangian density as well as the field equations above are well-defined for invertible as well as non-invertible vielbein. The solutions for the invertible vielbein are standard and well-known in the literature. For instance, under a suitable compactification, the theory is equivalent to the Kaluza-Klein framework \cite{kaluza,klein}. The other set of solutions for non-invertible vielbein were first explored in Ref.\cite{sengupta}, following which we provide a brief outline of the resulting effective four-dimensional theory.

Choosing the zero eigenvalue of the vielbein $\hat{e}_\mu^I$ to  lie along the fifth dimension ($v$):
\begin{eqnarray*}
\hat{e}_v^I=0,
\end{eqnarray*}
we may rewrite the vielbein as:
\begin{eqnarray}\label{e}
\hat{e}_\mu^I =
\left[\begin{array}{cc}
\hat{e}_a^i\equiv e_a^i & 0 \\
0 & 0 \\
\end{array}\right]
\end{eqnarray}
The world and internal indices are $\mu\equiv (a,v)\equiv (t,x,y,z,v)$ and $I\equiv (i,4)=(0,1,2,3,4)$, respectively.
The emergent tetrad fields $e_a^i$ (invertible) define the effective four-dimensional spacetime.  Their inverse are denoted as $e^a_i$ (not the same as $\hat{e}^a_i$ which do not exist):
\begin{eqnarray*}
e_a^i e^b_i=\delta_a^b,~e_a^i e^a_j=\delta _j^i.
\end{eqnarray*}
In terms of these, the corresponding $4$-metric is defined as $g_{ab}=e_a^i e_{bi}=\hat{g}_{ab}$. The extra dimensional components of the five-metric are manifestly trivial: $(\hat{g}_{va}=\hat{g}_{av}=0=\hat{g}_{vv})$. The four dimensional epsilon symbols, derived from the five dimensional antisymmetric tensor densities, are defined as: $\epsilon^{vabcd}\equiv \epsilon^{abcd},~\epsilon_{4ijkl}\equiv\epsilon_{ijkl}$.

 The most general set of solutions to the equations of motion (\ref{eom}) is then given by (further details regarding the method of solving the equations are available in \cite{sengupta}):
\begin{eqnarray}
\hat{w}_v^{~ij}=0=\hat{w}_v^{~4i},~\hat{w}_a^{4i}=e_a^j M^{ij}\equiv M_a^i;~\hat{w}_a^{~ij}=\bar{w}_a^{~ij}(e)+K_a^{~ij}
\end{eqnarray}
where $M^{ij}=M^{ji}$ is a symmetric matrix, $K_a^{~ij}=-K_a^{~ji}$ is contortion and $\bar{w}_a^{~ij}(e)=\frac{1}{2}[e^b_i\del_{[a}e_{b]}^j
-e^b_j\del_{[a}e_{b]}^i -  e_a^l e^b_i e^c_j
\del_{[b}e_{c]}^l]$  is the four-dimensional connection without torsion.
These are constrained by the equations of motion  as:
\begin{eqnarray}\label{kconstraint}
&& e^a_j K_a^{~ij}\equiv K_a^{~ia}=0,\nonumber\\
&& \left[\delta^a_b \delta_{kl}-e^a_k e_{bl} \right] D_a(\bar{w}) M^{kl}=0,\nonumber\\
&& e^a_i e^b_j \hat{R}_{ab}^{~~ij}= 0=e^a_i e^b_j \Big[\bar{R}_{ab}^{~~ij}+D_{[a}K_{b]}^{~ij}+K_{[a}^{~il}K_{b]l}^{~~~j}+M_{[a}^{i}M_{b]}^{j}\Big]
\end{eqnarray}
In the above, the field-strength $\bar{R}_{ab}^{~~ij}$ is defined with respect to the torsion-free connection $\bar{w}_a^{~ij}(e)$.
Here, we shall consider the simplest case, where both sets of emergent fields vanish:
\begin{eqnarray}\label{conditions}
K_a^{~ij}=0,~M^{kl}=0=M_a^i
\end{eqnarray} 
Note that the first condition implies a vanishing torsion in the four dimensional emergent spacetime.
With the above, the first two equations in (\ref{kconstraint}) reduce to identities, whereas the field equation for the tetrad reduces to a scalar constraint given by:
\begin{eqnarray}
e^a_i e^b_j \bar{R}_{ab}^{~~ij}~\equiv~ \bar{R}~=~0
\end{eqnarray}
Here $\bar{R}$ is defined as the four dimensional Ricci scalar without torsion.
This equation has the following general solution:
\begin{eqnarray}\label{EE-eff}
\bar{R}_{ab}-\frac{1}{2}g_{ab}\bar{R}~=~t_{ab}
\end{eqnarray}
where $t_{ab}$ is an arbitrary symmetric traceless tensor. Eq.(\ref{EE-eff}) represents the effective four dimensional equation of motion emerging from the 5D theory (\ref{L5}).

In the above, $t_{ab}$ plays the role of an emergent traceless energy-momentum tensor in four dimensions, whose origin is the five-dimensional geometry. Let us define the associated diagonal components of this tensor in a spherically symmetric (emergent) spacetime as: $[\rho^{(eff)},~P^{(eff)}_r,~P^{(eff)}_\theta,~P^{(eff)}_{\phi}=P^{(eff)}_\theta]$. The constraint $g^{ab}t_{ab}=0$ then translates to:
\begin{eqnarray}
\rho^{(eff)}-P^{(eff)}_r-2P^{(eff)}_\theta=0
\end{eqnarray}
Equivalently, the spatially averaged effective equation of state for this emergent 4D spacetime is given by: $w^{eff}=\frac{1}{3}$. Thus, the solutions  obtained earlier (\ref{f-radn}) for a non-ideal generalization of radiation fluid is a solution of the equations of motion (\ref{EE-eff}) as well. 

This concludes our demonstration that gravitational effects emerging from an extra dimension of vanishing proper length (in its most minimal form under the simplifying assumptions (\ref{conditions}) corresponds to a special case of the general solutions with flat rotation curves found for any $w$.

Before concluding this section, let us briefly remark upon another extra dimensional scenario in this context, namely, the Braneworld formulation \cite{rs,*rs1}. While it is quite different in essence and much less economical in its construction and structure than
the formulation discussed above, it could nevertheless reflect a
 similar connection to the galactic solutions here. To be precise, it is well-known that the 4D effective Einstein equations in this theory reflects a nonlocal geometric contribution (due to Weyl stresses) to the effective energy-momentum tensor \cite{shiromizu}. These encode appropriate projections of the five-dimensional Weyl tensor on the three-brane, leading to a traceless piece (denoted as $E_{\mu\nu}$ in Ref.\cite{shiromizu}) that is arbitrary upto the Bianchi identity. Such a term is non-vanishing unless the bulk 5D spacetime is not pure anti-de Sitter. We observe that if the energy momentum tensor generated by genuine matter along with the effective cosmological constant induced at the brane are ignored, the effective equation acquires a form similar to (\ref{EE-eff}) at the brane. Thus, within the braneworld framework as well, the non-ideal radiation geometries (\ref{f-radn}) emerge as exact solutions to the field equations.

\section{Conclusions}

Here we present a new set of galactic metrics for spirals. These are solutions to Einstein's equation with a stress-tensor, corresponding to anisotropic pressure and valid for any (spatially averaged) equation of state $w>-\frac{1}{3}$. The validity of these solutions is not necessarily restricted to the flat rotation curve region only, in contrast with the some of the geometries presented in the earlier literature \cite{nucamendi}. Apart from solutions associated with non-ideal `dust' and `radiation' fluid, the well-known Einstein cluster is shown to emerge as a special class whose EOS is determined purely by the limiting velocity of the asymptotic rotation curve.

The slope of the rotation curve solutions here reflect a gentle decline upto asymptotic distances for  sufficiently luminous spirals. This feature seems to be consistent with current observations. As another important result of direct observational relevance, we find the explicit value of the non-baryonic contribution to the gravitational bending of light for a large halo. The leading correction is shown to depend on $w$, which allows a concrete basis for a comparison with various `dark matter' models based on an assumed density profile. In other words, a precise experimental measurement of this correction could be used to determine the averaged EOS parameter of the anisotropic `dark matter' fluid.

Finally, we also show that the equations of motion elucidated here could also be generated through pure geometry under suitable assumptions. In particular, the solutions are relevant from the extra dimensional perspective. Here this connection has been explicitly illustrated for a candidate five-dimensional theory of gravity where the extra dimension has vanishing proper length. This is possible since this formulation corresponds to an weaker set of equations of motion compared to Einstein’s
theory. We observe that the same feature holds for the Braneworld scenario as well.

Among the many inviting challenges, it would be worthwhile  to apply our framework to the case of galaxy clusters. Since the X-ray gas is the dominant component of visible matter, the basic framework given by (\ref{EE}) continues to be valid in this case, provided an additional contribution from this gas is included at the right hand side. Given a density profile associated with this gas, the associated contribution to the energy momentum tensor could be determined assuming hydrostatic equilibrium. This exercise could be taken up elsewhere in detail.
Another important problem is to understand the empirical correlations between baryons and non-baryons in galaxies based on the new geometries presented here. This could signify an important test regarding the physical relevance of these spacetimes. 

\acknowledgments 
This work is supported by the MATRICS grant MTR/2021/000008, SERB, 
Govt. of India. Helpful comments from  Somnath Bharadwaj, Sayan Kar and Nemani V. Suryanarayana are gratefully acknowledged.

 \bibliography{FRC-NPB}
 
 \appendix
 
 \section{Determination of integration constants}
 
 In principle, the integration constants $C_{1,2}$ could be determined by matching the galactic halo metric to another metric at its boundary. This outer metric depends on the problem of interest in general. To be specific, here we demonstrate the procedure by assuming that the metric outside the boundary is given by Schwarzschild (asymptotically flat) and the one inside corresponds to the non-ideal `dust' ($w=0$) solution (\ref{w0}). The procedure to determine the constants for any other equation of state is exactly the same.
 
 The explained above, the halo metric (\ref{w0}) defines the spacetime at $r\leq R_0$ (upto some inner boundary).  Ignoring the higher order (in $\alpha$) corrections, this metric simplifies to:
 \begin{eqnarray}
 f_-(r)=\Big(\frac{r}{R}\Big)^{2\alpha} g_-^{-1}(r)\approx\frac{1}{(1+2\alpha+2\alpha^2)}\Big(\frac{r}{R}\Big)^{2\alpha}-C_1 \Big(\frac{r}{R}\Big)^{-1+\sqrt{\alpha}}+C_2 \Big(\frac{r}{R}\Big)^{-1-\sqrt{\alpha}}
  \end{eqnarray}
  The metric at $r\geq R_0$ is given by:
 \begin{eqnarray}\label{schw}
 f_+(r)=g_+^{-1}(r)=1-\frac{2M}{r}
 \end{eqnarray}
  where $M$ is the total mass enclosed within the radius $R_0$.
 Continuity of the metric at $r=R_0$ implies:
 \begin{eqnarray}
 R_0=R,~C_1-C_2=\frac{2M}{R_0}-\frac{2\alpha(1+\alpha)}{1+2\alpha+2\alpha^2}
 \end{eqnarray}
 Note that the scale $R$ in the solutions define the radius of the halo. 
 Further, the derivatives of the metric are also continuous provided:
 \begin{eqnarray}\label{c+}
 C_1+C_2=-\frac{2\alpha^{\frac{3}{2}}}{1+2\alpha+2\alpha^2}
 \end{eqnarray}
 Ignoring $o(\alpha^2)$ corrections, the resulting solutions finally read:
 \begin{eqnarray}
 C_1\approx \frac{M}{R_0}-\frac{\alpha(1+\sqrt{\alpha})}{1+2\alpha},~C_2\approx -\frac{M}{R_0}+\frac{\alpha(1-\sqrt{\alpha})}{1+2\alpha}
 \end{eqnarray}
 
\section{Energy conditions} 
 
In this section, we analyze the energy conditions in the context of the galactic spacetime solutions obtained earlier. 
As in the previous section, we shall analyze the non-ideal `dust' case as an illustrative example.

The density and pressure components for $w=0$ as given by eq.(\ref{pw0}) could be rewritten as follows:
\begin{eqnarray}\label{pw1}
8\pi\rho &~\approx~&\frac{1}{r^2}\Bigg[\frac{2\alpha(1+\alpha)}{1+2\alpha+2\alpha^2}+C_1 \sqrt{\alpha}\left(\frac{r}{R}\right)^{-1+\sqrt{\alpha}}+C_2 \sqrt{\alpha}\left(\frac{r}{R}\right)^{-1-\sqrt{\alpha}}\Bigg],\nonumber\\
8\pi P_r &~\approx~&-\frac{1}{r^2}\Bigg[\frac{2\alpha^2}{1+2\alpha+2\alpha^2}+C_1 \sqrt{\alpha}\left(\frac{r}{R}\right)^{-1+\sqrt{\alpha}}+C_2 \sqrt{\alpha}\left(\frac{r}{R}\right)^{-1-\sqrt{\alpha}}\Bigg],\nonumber\\
8\pi P_\theta &~\approx~&\frac{1}{r^2}\Bigg[\frac{\alpha^2}{1+2\alpha+2\alpha^2}+\frac{C_1}{2} \sqrt{\alpha}\left(\frac{r}{R}\right)^{-1+\sqrt{\alpha}}+\frac{C_2}{2} \sqrt{\alpha}\left(\frac{r}{R}\right)^{-1-\sqrt{\alpha}}\Bigg]
\end{eqnarray}
 In the above, we have kept only the leading order coefficients in the terms involving $C_{1,2}$, since that is sufficient in the weak-field limit. These lead to the following identities at the leading order:
 \begin{eqnarray}\label{pw1}
8\pi(\rho+P_r) &~\approx~&\frac{2\alpha}{(1+2\alpha)r^2},\nonumber\\
8\pi (\rho+P_\theta) &~\approx~&-\frac{1}{r^2}\Bigg[\frac{2\alpha}{1+2\alpha}+\frac{3 \sqrt{\alpha}(C_1+C_2)R}{r}\Bigg],\nonumber\\
8\pi (\rho+\sum_i P_i) &~\approx~& \frac{1}{r^2}\Bigg[\frac{2\alpha}{1+2\alpha}+\frac{ \sqrt{\alpha}(C_1+C_2)R}{r}\Bigg],\nonumber\\
8\pi (\rho-P_r) &~\approx~&\frac{1}{r^2}\Bigg[\frac{2\alpha}{1+2\alpha}+\frac{2 \sqrt{\alpha}(C_1+C_2)R}{r}\Bigg],\nonumber\\
8\pi (\rho-P_\theta) &~\approx~&\frac{1}{r^2}\Bigg[\frac{2\alpha}{1+2\alpha}+\frac{ \sqrt{\alpha}(C_1+C_2)R}{2r}
\Bigg]
\end{eqnarray}
Upon using eq.(\ref{c+}) obtained at the last section, it is straightforward to see that the weak ($\rho\geq  0,~\rho+P_i\geq 0$), null ($\rho+P_i\geq 0$), strong ($\rho+P_i\geq 0,~\rho+\sum_i P_i\geq 0$) and dominant ($\rho\geq 0,~\rho\geq |P_i|$) energy conditions are satisfied, provided:
\begin{eqnarray}
r\geq 2\alpha R_0,
\end{eqnarray}
$R_0$ being the halo radius. The critical radius at the right hand side above is very small. Since this radius is practically much smaller than the typical inner radii where luminous matter at the interior starts dominating, the solution (\ref{pw0}) satisfies the energy conditions everywhere as long as the region of interest lies sufficiently far away from luminous matter (as assumed in this work).
 
 \end{document}